\documentclass[11pt]{article}
\usepackage{verbatim}
\usepackage{amsfonts}
\usepackage{amsbsy}
\usepackage{epsfig}
\usepackage{latexsym}
\newcommand{\qed}{\hfill $\Box$ \medskip}

\makeatletter \@addtoreset{equation}{section} \makeatother

\title{{\Large \bf{Ricci Solitons and Einstein-Scalar Field
Theory}}}
\author{\\M M Akbar\footnote
{E-mail: makbar@math.ualberta.ca}
 and E Woolgar\footnote
{E-mail: ewoolgar@math.ualberta.ca}
\\
Department of Mathematical and Statistical Statistics,
\\ University of Alberta, \\ Edmonton, Alberta\\
Canada T6G 2G1}
\date{\today}

\begin{document}
\maketitle

\begin{abstract}
\noindent B List has recently studied a geometric flow whose fixed
points correspond to static Ricci flat spacetimes. It is now known
that this flow is in fact Ricci flow modulo pullback by a certain
diffeomorphism. We use this observation to associate to each static
Ricci flat spacetime a local Ricci soliton in one higher dimension.
As well, solutions of Euclidean-signature Einstein gravity coupled
to a free massless scalar field with nonzero cosmological constant
are associated to shrinking or expanding Ricci solitons. We exhibit
examples, including an explicit family of complete expanding
solitons. These solitons can also be thought of as a Ricci flow for
a complete Lorentzian metric. The possible generalization to
Ricci-flat stationary metrics leads us to consider an alternative to
Ricci flow.
\end{abstract}

\newpage

\section{Introduction}

\noindent Geometric flows have become important tools in Riemannian
geometry and general relativity. Quite early on, Geroch
\cite{Geroch} introduced the inverse mean curvature flow in an
argument in support of the conjecture that later became the positive
mass theorem. The method eventually led to the Huisken-Ilmanen proof
of the Riemannian Penrose conjecture \cite{HI}. More recently, the
powerful tool of Ricci flow has been used to prove the Poincar\'e
conjecture \cite{MT} and, it appears, the Thurston conjecture as
well \cite{CZ,MT2}. This flow is given as
\begin{equation}
\frac{\partial g_{\mu\nu}}{\partial \lambda} = -2R_{\mu\nu} \ .
\label{eq1.1}
\end{equation}
The flow is often generalized by pulling back along an evolving
(i.e., $\lambda$-dependent) diffeomorphism. This yields the
Hamilton-DeTurck flow
\begin{equation}
\frac{\partial g_{\mu\nu}}{\partial \lambda} = -2R_{\mu\nu} +
\pounds_X g_{\mu\nu}\ , \label{eq1.2}
\end{equation}
where $X$ is the vector field generating the diffeomorphism.

The power of geometric flows derives in part from the
(quasi-)parabolic character of the flow equations on Riemannian
geometries. This property is typically not present for flows on
pseudo-Riemannian geometries (spacetimes), unless the problem can be
phrased in Riemannian terms, as is the case for the Riemannian
Penrose conjecture. Following this reasoning, one may expect
quasi-parabolicity of the Ricci flow on a spacetime if the metric
has a static or perhaps even a stationary Killing vector field
(recall that the Ricci flow preserves isometries).\footnote
{Parabolic flows sometimes arise even in nonstationary spacetimes.
Consider the interesting example of Robinson-Trautman 4-metrics, for
which the Einstein equation becomes the fourth-order Calabi flow in
2 dimensions \cite{Tod, PTC}.}

A geometric flow of static Lorentzian metrics was studied by List
\cite{List}. He did not begin with the Ricci flow of a static metric
(i.e., a metric with timelike, hypersurface-orthogonal Killing
field). Instead he presented a system of flow equations whose fixed
points solve the static vacuum Einstein equations, but which seemed
better suited to obtaining estimates than Ricci flow. List's system
is
\begin{eqnarray}
\frac{\partial g_{ij}}{\partial \lambda} &=& -2 \left ( R_{ij}
-k_n^2 \nabla_i u \nabla_j u \right )\ , \label{eq1.3}\\
\frac{\partial u}{\partial \lambda} &=& \Delta u\ , \label{eq1.4}
\end{eqnarray}
where, for each value of the flow parameter $\lambda$,
$g_{ij}(\lambda ;x)$ is a Riemannian metric on an $n$-manifold,
$u(\lambda;x)$ is a function, $\Delta u := g^{ij}\nabla_i\nabla_j u$
is the Laplacian of $u$, $k_n$ is an arbitrary constant (which we
will set to $\sqrt{\frac{n-1}{n-2}}$ when $n>2$), and $-e^{2u} dt^2
+e^{\frac{2u}{n-2}}g_{ij}$ is then the (flowing) spacetime static
metric.

Static metrics have always yielded important examples of exact solutions in general relativity. They also arise, for example, as the result of limiting processes along a family of metrics. This occurs in two contexts that we know of: Anderson's approach to geometrization of 3-manifolds \cite{Anderson1, Anderson2}, and Bartnik's approach to quasi-local mass \cite{Bartnik}. In fact, the motivation behind List's flow was to have a method of generating families of metrics that might be well-suited to studying questions arising from Bartnik's mass definition \cite{Huisken}. In the special case of a bounded region $B$ in a time-symmetric slice of spacetime, Bartnik defines the quasilocal mass of $B$ by considering all asymptotically flat Riemannian manifolds having non-negative scalar curvature into which $B$ can be embedded (smoothly up to $\partial B$),\footnote
{At $\partial B$, ``geometric boundary conditions'' are imposed.
The induced metric and mean curvature must match across $\partial B$,
but the full extrinsic curvature of $\partial B$ need not.}
without there being a stable minimal sphere outside $B$. Each such
manifold has a non-negative ADM mass. Bartnik's quasilocal mass is
the infimum of these ADM masses, and is therefore non-negative as
well. What is not clear {\it a priori} is whether it is
positive.\footnote
{Indeed, the reason for the condition banning stable minimal spheres
outside $B$ is that manifolds which such spheres can have
arbitrarily small ADM masses, which would cause that the Bartnik
mass to be driven to zero; see \cite{Bartnik}.}
However, Bartnik has conjectured that the infimum in the definition
is realized by an extension of $B$ which solves the static vacuum
Einstein equations: by the positive mass theorem, then $m_B$ will be
nonzero unless $B$ embeds in flat space. The problem is then to show
that such a mass-minimizing extension exists.

Huisken and Ilmanen \cite{HI} have since proved that $m_B>0$ unless $B$
is isometric to a domain in flat space, but Bartnik's conjecture
remains open. One method of attacking it would be to try to
construct the conjectured static metric as a geometric limit of a
suitable flow starting from some initial data, say a positive scalar
curvature metric on the extension, and subject to boundary
conditions at $\partial B$. List's flow is a promising candidate
because, as well as having static metrics as fixed points, it
preserves asymptotic flatness, has compactness properties to make
sense of limits \cite{List}, and ``almost'' preserves positive
scalar curvature (along the flow, the scalar curvature is bounded
below by $-const/(1+\lambda)$, which tends to zero as
$\lambda\to\infty$).

It is now realized that List's system of flow equations is in fact
the pullback by a certain diffeomorphism of a class of Ricci flows
in one higher dimension; i.e., List's system is really a certain
Hamilton-DeTurck flow. This does not make List's flow any less
interesting however. An advantage of this is that Ricci flow results
can be employed to study List's flow. But as well, certain Ricci
flow results may be more evident from the perspective of List's
flow; i.e., by employing the diffeomorphism gauge that leads to
List's flow equations. The point of this paper is to document one
example of this interplay, arising from a search for simple solutions of List's equations.

Applications of geometric flow problems often require quite detailed
analytical arguments, for which it is best first to have intuition
developed from exact solutions. In the case of Ricci flow, the
simplest solutions are the fixed points, i.e., Ricci-flat metrics.
The next simplest are the Ricci solitons.

\bigskip
\noindent {\bf Definition 1.1.} A {\it Ricci soliton} is a
manifold-with-metric $(M^{n+1},g_{\mu\nu})$ and vector field $X$ on
it such that
\begin{equation}
R_{\mu\nu}-\frac12\pounds_X g_{\mu\nu}=\kappa g_{\mu\nu} \label{eq1.5}
\end{equation}
for some constant $\kappa$.
\bigskip

\noindent Here $R_{\mu\nu}$ is the Ricci curvature of $g_{\mu\nu}$.
The soliton is {\it gradient} if $X=\nabla\varphi$ for some function
$\varphi$ and {\it steady} if $\kappa=0$. If $\kappa<0$ the soliton
is called an {\it expander}; if $\kappa>0$ it is a {\it shrinker}.
Finally, a {\it local} soliton is one that solves (\ref{eq1.5}) on
an open region which might not admit an extension to a complete
manifold with the soliton metric.

Given a pair $(X,g)$ solving (\ref{eq1.5}), the metric $\kappa
\lambda \varphi^*(\lambda)g_{\mu\nu}$ obtained by pulling back
$g_{\mu\nu}$ along $\frac{1}{\kappa \lambda} X$ and rescaling by
$\kappa \lambda$, solves (\ref{eq1.1}) \cite{Chow1}. Solitons are
not fixed points of (\ref{eq1.1}) but evolve only by diffeomorphism
and scaling, and in this sense are the simplest nontrivial
solutions.

As we said earlier, List's flow admits fixed point solutions
corresponding to static spacetime metrics. The next simplest
solutions of List's flow are {\it solitons of List's flow}, which
are flows constructed from pairs $(u,g_{ij})$ that obey
\begin{eqnarray}
&&R_{ij}-k_n^2\nabla_i u \nabla_j u -\frac12 \pounds_X g_{ij}
= \kappa g_{ij}\ , \label{eq1.6}\\
&&\Delta u + \pounds_X u = 0 \ , \label{eq1.7}
\end{eqnarray}
for some constant $\kappa$ and vector field $X$, where now $R_{ij}$
is the Ricci curvature of $g_{ij}$. When $X$ vanishes, equations
(\ref{eq1.6}, \ref{eq1.7}) are well-known in general relativity:

\bigskip
\noindent {\bf Definition 1.2.} We will call the equations
\begin{eqnarray}
&&R_{ij}-k_n^2\nabla_i u \nabla_j u
= \kappa g_{ij}\ , \label{eq1.8}\\
&&\Delta u = 0 \ , \label{eq1.9}
\end{eqnarray}
the {\it Einstein free-scalar-field system}. When $\kappa=0$ as
well, equations (\ref{eq1.8}, \ref{eq1.9}) are called the {\it
static vacuum Einstein equations}.
\bigskip

The terminology {\it Einstein free-scalar-field} arises because, for
Lorentzian signature $g_{ij}$, equations (\ref{eq1.8}, \ref{eq1.9})
describe Einstein gravity with cosmological constant, coupled to a
free scalar field. But note that we will use the terminology without
regard to the signature of $g_{ij}$. The {\it static vacuum
Einstein} terminology arises because, if $g_{ij}$ has Euclidean
signature and when $\kappa=0$ and the conventional choice
$k_n=\sqrt{\frac{n-1}{n-2}}$ is made, equations (\ref{eq1.8},
\ref{eq1.9}) imply that the metric on ${\mathbb R}\times M^n$,
$n>2$, given by
\begin{equation}
dS^2=G_{\mu\nu}dx^{\mu}dx^{\nu}=-e^{2u}dt^2+e^{-\frac{2u}{n-2}}g_{ij}dx^idx^j
\label{eq1.10}
\end{equation}
is Ricci-flat and $\frac{\partial}{\partial t}$ is a
hypersurface-orthogonal Killing vector field.

In fact, equation (\ref{eq1.9}) is redundant for any
$\kappa\in{\mathbb R}$ because it can be derived from (\ref{eq1.8})
using the contracted second Bianchi identity. Thus it plays only the
role of an integrability condition for (\ref{eq1.8}).

The main result of this paper is that the observation that List's
flow is a Hamilton-DeTurck flow leads directly to a nice connection
between Ricci solitons and the Einstein free-scalar-field system:

\bigskip
\noindent {\bf Observation 1.3.} {\sl Solutions the Einstein
free-scalar-field system correspond to Ricci solitons.}
\bigskip

\noindent Indeed, any soliton of List's flow (i.e., any solution of
(\ref{eq1.6}, \ref{eq1.7})) corresponds to a Ricci soliton, but we
will focus on the relation to relativity theory expressed in the
Observation.

We give the precise correspondence in section 2. Many of the
solitons that arise in this manner are only local solitons, in the
sense that the metric is not complete. This is not unexpected in
view of various theorems in physics for static vacuum Einstein and
Einstein-scalar systems, such as the theorem of Lichnerowicz
\cite{Lich}, ``no hair'' theorems, and singularity theorems (see,
e.g., Chase \cite{Chase}).

In section 3, we identify the local solitons that arise from several
familiar Einstein-scalar solutions. We also construct an example of
a complete soliton arising from the Einstein-scalar system with
negative cosmological constant. This may also be thought of as a
nontrivial Ricci flow on a complete Lorentzian manifold.

In section 4 we generalize from static to stationary metrics.
Stationary Lorentzian metrics have a timelike Killing vector field,
but are more general than static metrics because this vector field
is no longer necessarily hypersurface-orthogonal; when it is not, we
say the metric is {\it rotating}. We ask whether there is a Ricci
flow adapted to stationary metrics in the way that List's flow is
adapted to static metrics; namely, are there metric-diffeomorphism
pairs satisfying (\ref{eq1.2}) such that the fixed points yield
Ricci-flat stationary rotating metrics? If so, then our soliton
construction would extend to that case. However, we find that for
the most obvious choice of diffeomorphism at least, the fixed points
of the resulting Hamilton-Deturck flow equations do not coincide
with rotating Ricci-flat metrics. This leads us to propose an
alternative flow, for which fixed points are Ricci-flat stationary
metrics, but which we do not derive from Hamilton-DeTurck flow.

\section{Solitons and Free Scalar Fields}
\setcounter{equation}{0}

\subsection{The precise relationship}

\noindent The precise form of Observation 1.3 is the following result:

\bigskip \noindent
{\bf Proposition 2.1.} {\sl If the pair $(u,g_{ij})$ solves
(\ref{eq1.8}) (and thus (\ref{eq1.9}) as well), the metric
\begin{equation}
ds^2 = g_{\mu\nu}dx^{\mu}dx^{\nu}=e^{2k_n u}dt^2+g_{ij}dx^idx^j
\ . \label{eq2.1}
\end{equation}
is a local Ricci soliton on ${\mathbb R}\times M^n$ solving
(\ref{eq1.5}).}
\bigskip

\noindent {\bf Proof.} The Ricci curvature of the metric
(\ref{eq2.1}) is
\begin{eqnarray}
R_{\mu\nu} dx^{\mu}dx^{\nu}&=&- e^{2k_nu} \left ( k_n \Delta u + k_n^2
\vert\nabla u\vert^2 \right ) dt^2\nonumber\\ && + \left (
R^g_{ij}-k_n^2\nabla_iu\nabla_ju-k_n\nabla_i\nabla_ju\right ) dx^i dx^j
\nonumber\\
&=&- k_n^2 e^{2k_nu} \vert \nabla u \vert^2 dt^2+ \left (
\kappa g_{ij}-k_n\nabla_i\nabla_ju\right ) dx^i dx^j \ ,
\label{eq2.2}
\end{eqnarray}
where we used (\ref{eq1.8}, \ref{eq1.9}) and the shorthand $|\nabla
u |^2:=g^{ij}\nabla_i u \nabla_j u$. On the other hand, define the
vector field
\begin{equation}
X:=-\kappa t\frac{\partial}{\partial t} -k_n g^{ij}\nabla_i u
\frac{\partial}{\partial x^j} \ . \label{eq2.3}
\end{equation}
Using the metric (\ref{eq2.1}), it's straightforward to compute that
\begin{equation}
\nabla_{\mu}X_{\nu}= \left [ \begin{array}{cc} -e^{2k_nu}\left (
k_n^2 \vert \nabla u \vert^2+\kappa \right ) & \kappa k_n e^{2k_nu}
\nabla_i u\\ - \kappa k_n e^{2k_nu} \nabla_i u & -k_n\nabla_i\nabla_j u
\end{array} \right ] \ . \label{eq2.4}
\end{equation}
We note in passing that since this is not symmetric, $X$ is not a
gradient vector field, but $X\wedge dX=0$ so $X$ is
hypersurface-orthogonal (here we have used $X$ to denote both the
vector field and its metric-dual 1-form). From (\ref{eq2.4}) we get
that
\begin{equation}
\pounds_X g_{\mu\nu} =\nabla_{\mu}X_{\nu}+\nabla_{\nu}X_{\mu} =\left
[ \begin{array}{cc} -2e^{2k_nu}\left ( k_n^2\vert \nabla u \vert^2+\kappa
\right ) & 0\\ 0 & -2k_n\nabla_i\nabla_j u \end{array} \right ]\ .
\label{eq2.5}
\end{equation}
Combining (\ref{eq2.2}) and (\ref{eq2.5}) yields (\ref{eq1.5}).
\hfill \qed
\bigskip

\noindent{\bf Remark 2.2.} From (\ref{eq1.5}), for each sign of
$\kappa$, there are constant $u$ solitons $dt^2+g_{ij}dx^idx^j$ with
$g_{ij}$ an Einstein metric. If $\kappa=0$, these exist for $t\in
S^1$ as well as for $t\in{\mathbb R}$, but for $\kappa\neq 0$, the
vector field $X$ would not be single-valued if $t$ were periodic.
\bigskip

\noindent{\bf Remark 2.3.} The invariance $u\to -u$ of equation
(\ref{eq1.8}) yields a second, distinct soliton by replacing $u$ by
$-u$ in (2.4). This is an example of the Buscher duality described
in \cite{Chow1}.
\bigskip

\noindent{\bf Remark 2.4.} We can replace (\ref{eq2.1}) by
\begin{equation}
ds^2 = -e^{2k_n u}dt^2+g_{ij}dx^idx^j\ , \label{eq2.6}
\end{equation}
thereby obtaining a Ricci soliton of Lorentzian signature.
\bigskip

\noindent{\bf Remark 2.5.} It is common to represent scaling
solitons in $\lambda$-dependent form. In the present case, two such
forms are
\begin{equation}
ds^2 = G_{\mu\nu}dx^{\mu}dx^{\nu}=-2\kappa \lambda \left ( \pm
e^{2k_n u}dt^2 + g_{ij}dx^idx^j \right ) \ , \label{eq2.7}
\end{equation}
and
\begin{equation}
ds^2 = G_{\mu\nu}dx^{\mu}dx^{\nu} = \pm e^{2k_nu}dt^2 -
2\kappa \lambda g_{ij}dx^idx^j \ , \label{eq2.8}
\end{equation}
where we take $\lambda\in (-\infty,0)$ if $\kappa>0$ for the
shrinker and $\lambda\in (0,\infty)$ if $\kappa<0$ for the expander.
These metrics solve the Hamilton-DeTurck flow equation
\begin{equation}
\frac{\partial G_{\mu\nu}}{\partial \lambda} =
-2R^G_{\mu\nu}+\pounds_Y G_{\mu\nu} \label{eq2.9}
\end{equation}
with vector fields $Y=-\frac{1}{2\kappa\lambda}X$ and
$Y=\frac{k_n}{2\kappa\lambda}g^{ij}\nabla_i u
\frac{\partial}{\partial x^j}$, respectively.

\bigskip

\subsection{Completeness}

\noindent The question arises as to when the solitons (\ref{eq2.1})
are complete. Some conclusions can be drawn from simple properties
of global solutions of (\ref{eq1.8}) and (\ref{eq1.9}):

\bigskip

\noindent{\bf Lemma 2.6.} {\sl If $(M^n,g)$ is a closed manifold, then
it is Einstein, $u=const$, and the soliton $dt^2+g_{ij}$ is
complete. If $(M^n,g)$ is noncompact and complete, then $\kappa\le
0$. If, further, $\kappa=0$ and if $\vert \nabla u \vert \to 0$ at
infinity, then $u=const$ and $(M,g)$ and the soliton are Ricci-flat.}
\bigskip

\noindent{\bf Remark 2.7.} If $n=3$ and $\kappa=0$, Anderson's generalization \cite{Anderson2} of Lichnerowicz's theorem \cite{Lich} shows that the soliton will be flat space, even without any assumption on $|\nabla u|$.
\bigskip

\noindent{\bf Proof of 2.6.} The first statement of Lemma 2.6 follows from
(\ref{eq1.9}) and the strong maximum principle. The second statement
follows from (\ref{eq1.8}) and Myers' diameter estimate. To prove
the third statement, set $\kappa=0$ in (\ref{eq1.8}) and use it and
(\ref{eq1.9}) to compute that
\begin{equation}
\Delta \left ( \vert \nabla u \vert^2 \right )
= 2k_n^2\left ( \vert \nabla u \vert^2
\right )^2+2\vert \nabla \nabla u \vert^2 \ge 0\ . \label{eq2.10}
\end{equation}
Since $M$ is complete and $u\in C^2(M)$ is globally defined and
$|\nabla u| \to 0$ at infinity, it then follows from the strong maximum
principle that $u$ is constant.\footnote
{That is, since $\vert \nabla u\vert^2$ tends to zero at infinity,
it must achieve a maximum in $M$. At a maximum point, the left-hand
side of (\ref{eq2.10}) would be $\le 0$ but the right-hand side
would be $\ge 0$, so both sides must be zero. This forces $\vert
\nabla u\vert^2=0$ to be the maximum, so $\vert \nabla u\vert^2$
vanishes pointwise. $u=const$.}
\hfill\qed
\bigskip

Incompleteness can arise because of a singularity---i.e., inextenbility of (\ref{eq2.1}) in any coordinates---or it can also arise merely because the coordinate system breaks down at fixed points of the Killing field
$\frac{\partial}{\partial t}$.
The latter case is analogous to the incompleteness in the
Schwarzschild exterior black hole metric after Wick rotation to
Riemannian signature, which is cured by adding in the fixed point at
$r=2m$. Lemma 2.6 does not distinguish between these sources of incompleteness. The following lemma shows that the soliton will have a singularity when the Einstein-scalar system does. This is useful because, for $g_{ij}$ a Lorentzian metric, the Einstein-scalar system has been studied in the context of ``no hair '' results, so we can Wick rotate the conclusions to our case. These results show that, in some circumstances at least, extension of (\ref{eq2.1}) to a fixed point of the Killing field is not possible. We will return to this point
later (see footnote \ref{fn5}).

\bigskip
\noindent{\bf Lemma 2.8.} {\sl The soliton is inextendible wherever
the norm of either ${\rm sec}[g]$, $|\nabla u|$, or $|\nabla\nabla
u|$ diverges.}
\bigskip

\noindent Here ${\rm sec}[g]$ refers to the sectional curvatures of
$g$; i.e., certain orthonormal components of ${\rm Riem}[g]$ that
fully determine ${\rm Riem}[g]$.
\bigskip

\noindent{\bf Proof.}
We compute scalar invariants of the soliton metric (\ref{eq2.1}) to get
\begin{eqnarray}
R&=&-k_n^2 \vert \nabla u \vert^2+n\kappa = -R^g +2n\kappa\ ,
\label{eq2.11}\\
R_{\mu\nu}R^{\mu\nu}&=&k_n^4 \left ( \vert \nabla u \vert^2 \right )^2
+n\kappa^2+k_n^2\vert \nabla \nabla u\vert^2 \label{eq2.12}\\
R_{\alpha\beta\mu\nu}R^{\alpha\beta\mu\nu}&=&4k_n^2\vert \nabla \nabla u
+ k_n \nabla u \nabla u \vert^2 + \vert {\rm Riem}[g]\vert^2 \ .
\label{eq2.13}
\end{eqnarray}
Curvatures on the left-hand side refer to the metric (\ref{eq2.1}).
On the right-hand side, $\vert \cdot \vert$ is the norm with respect
to $g^{ij}$ and ${\rm Riem}[g]$ is the Riemann tensor of $g_{ij}$.
Then the result follows by inspection of these equations. \hfill\qed
\bigskip

In the remainder of this section, we briefly discuss a case left
open by Lemma 2.6. Consider $\kappa=0$ and $\vert \nabla u\vert \to
c>0$ or $\vert \nabla u\vert \to \infty$. Assume further that there
is a unit speed, geodesic ray\footnote
{A ray minimizes arclength between any two of its points.}
$\gamma: [0,\infty) \to M$ along which
\begin{equation}
{\dot \gamma}\cdot \nabla u = | \nabla u | \cos \theta_0 \ge
\epsilon \cos \theta_0 =: C>0\ . \label{eq2.14}
\end{equation}
This condition always holds for $(M,g_{ij})$ asymptotically flat
with $u\sim const/r$ at large $r$. In such a case, we can foliate
the asymptotic region by convex level sets of $u$ and choose
$\gamma$ such that ${\dot \gamma}$ makes an angle with $\nabla u$
that is never greater than some $\theta_0<\frac{\pi}{2}$.

\bigskip

\noindent{\bf Lemma 2.9.} {\sl There are no solutions of (\ref{eq1.8})
for which (\ref{eq2.14}) holds along a ray $\gamma:[0,\infty)\to M$,
and thus no asymptotically flat solitons (\ref{eq2.1}) with $|
\nabla u | \to c \in (0,\infty]$.}

\bigskip

\noindent The argument relies on properties of (\ref{eq1.8}) but not
on the soliton interpretation, and is a simple variation on the
textbook proof of Myers' diameter estimate using Synge's formula.
Contrary to Myers' estimate, (\ref{eq1.8}) implies that only one
eigenvalue of Ricci is nonzero, but this is sufficient in the given
circumstances.

\bigskip

\noindent{\bf Proof.} In the standard way construct an
orthonormal basis $\{ {\dot \gamma}(t), e_{(i)}(t)|i=2,\dots, n \}$
along $\gamma$ and use it to define $(n-1)$ orthogonal variation
vector fields that vanish at $\gamma$'s endpoints:
\begin{equation}v_{(i)}:=f(t)e_{(i)}=\sin\left ( \frac{\pi t}{b-a}
\right ) e_{(i)} \ . \label{eq2.15}
\end{equation}

Synge's formula \cite{Petersen} for the second variation of
$\gamma$'s energy along $v_{(i)}$ reads
\begin{equation}
E_{(i)}''=\int\limits_a^b \left [ g(\nabla_{\dot \gamma} v_{(i)},
\nabla_{\dot \gamma}v_{(i)})-g(R(v_{(i)},
{\dot \gamma}){\dot \gamma},v_{(i)} \right ] dt \ . \label{eq2.16}
\end{equation}
In (\ref{eq2.16}), $i$ labels distinct vector fields and so there is
no sum over it, but if we do sum over $i\in \{ 2, \dots, n \}$ we obtain
\begin{eqnarray}
\sum\limits_{i=2}^n E_{(i)}''&=&\int\limits_a^b \left [
(n-1){\dot f}^2(t)-f^2(t)R_{ij}
{\dot \gamma}^i{\dot \gamma}^j \right ] dt \nonumber\\
&=&\int\limits_a^b \left [ (n-1){\dot f}^2(t)-k_n^2 f^2(t)
\left ( {\dot \gamma}^i\nabla_i u\right )^2 \right ] dt \nonumber\\
&\le&\int\limits_a^b \left [ \frac{(n-1)\pi^2}{(b-a)^2}
\cos^2\left ( \frac{\pi t}{b-a}\right ) -C^2 k_n^2 \sin^2\left (
\frac{\pi t}{b-a}\right ) \right ] dt \nonumber\\
&=&\frac{(n-1)\pi^2}{2(b-a)}-\frac12 C^2 k_n^2 (b-a) \ , \label{eq2.17}
\end{eqnarray}
where we have used (\ref{eq1.8}), (\ref{eq2.15}), and the condition
on $\gamma\cdot \nabla u$. For $(b-a)$ large enough, the right-hand
side of (\ref{eq2.17}) is less than zero, and then there is an $i$
such that $E_{(i)}''<0$. By standard results, there is then a
shorter geodesic than $\gamma$; i.e., $\gamma:[a,b]\to M$ cannot be
a ray. This is a contradiction. \qed
\bigskip

\section{Examples}

\subsection{Introduction}

\noindent In this section, we give examples of solitons constructed
from solutions of the Einstein-free scalar field system.

Perhaps one's first thought is to ask what soliton (or solitons)
arises in this manner from the Schwarzschild metric. Therefore we
develop in this subsection the specific form of equation
(\ref{eq1.8}) as it applies to metrics $g_{ij}$ of the form
\begin{equation}
ds^2=g_{ij}dx^idx^j=dr^2+f^2(r)g_{ab}dx^a dx^b =dr^2+f^2(r)d\Omega^2_k
\ , \label{eq3.1}\\
\end{equation}
where $d\Omega^2_k$ is now an Einstein metric with scalar curvature
$k$ normalized to $-1$, $0$, or $1$. This form includes the
Schwarschild ${\rm SO}(n-1)$-symmetric case but is somewhat more
general, allowing us to apply the equations developed here to most
other, but not entirely all, solitons that we discuss in subsequent
subsections.

Taking $u=u(r)$, then the integrability condition (\ref{eq1.9})
shows that
\begin{equation} u'(r)=\frac{A}{\left (f(r)\right )^{n-1}}\ ,
\qquad A=const\ . \label{eq3.2}
\end{equation}
We take $A\neq 0$ since the $u=const$ case was discussed in Remark
2.2. Note that $u(r)$ has no critical point in the interior of the
domain of $r$. This is consistent with the maximum principle.

The Ricci curvature of the metric $g_{ij}$ is
\begin{equation}
R_{ij}dx^idx^j = -(n-1)\frac{f''(r)}{f(r)}dr^2 + \left [ (n-2)\left
( k-f'^2(r)\right ) -f(r)f''(r)\right ] d\Omega^2_k \ .
\label{eq3.3}
\end{equation}
Then equation (\ref{eq1.8}) leads to the two equations
\begin{eqnarray}
\frac{f''(r)}{f(r)}&=&-\frac{1}{(n-1)} \left [ k_n^2 u'^2(r)+\kappa
\right ] =- \frac{A^2}{(n-2)f^{2(n-1)}(r)}-\frac{\kappa}{n-1}
\ , \label{eq3.4}\\
\frac{f''(r)}{f(r)}&=&(n-2)\frac{\left ( k-f'^2(r)\right )}{f^2(r)}
-\kappa \ , \label{eq3.5}
\end{eqnarray}
where we have used (\ref{eq3.2}). Eliminating $f''(r)$, we obtain
\begin{equation}
f'^2(r) + \frac{\kappa}{(n-1)}f^2(r)-\frac{k_n^2 A^2}{(n-1)(n-2)
f^{2n-4}(r)} =k \ . \label{eq3.6}
\end{equation}
Since (\ref{eq3.6}) may also be obtained by multiplying
(\ref{eq3.4}) by $f(r)f'(r)$ and integrating, the problem reduces to
solving (\ref{eq3.6}) and discarding any $f'(r)=0$ solutions that do
not solve (\ref{eq3.4}, \ref{eq3.5}).

Now (\ref{eq3.6}) has the familiar form of a unit mass with total
energy $k/2$ moving in a central potential
\begin{eqnarray}
&&\frac{1}{2}{\dot \rho}^2 + V(\rho)=k/2\ , \label{eq3.7}\\
&&V(\rho)=\frac{1}{2}\left (\frac{\kappa\rho^2}{n-1}
-\frac{A^2}{(n-2)^2\rho^{2n-4}}\right )\label{eq3.8}\ ,
\end{eqnarray}
where $\rho:=f(r)$, ${\dot \rho}:=f'(r)$, and we've used
$k_n^2=\frac{n-1}{n-2}$.

\subsection{$n=3$ dimensions}

For simplicity of the presentation, we will fix the dimension and
consider in this subsection only the $n=3$ case. Then
\begin{equation}
V(\rho)=\frac12 \left ( \frac{\kappa\rho^2}{2}-\frac{A^2}{\rho^2}
\right ) \ . \label{eq3.9}
\end{equation}

There are $\rho=\rho_0\equiv const>0$ solutions of (\ref{eq3.7},
\ref{eq3.9}) but some of these are spurious and do not solve
(\ref{eq3.4}, \ref{eq3.5}). There is, however, a genuine solution
with $\rho=\rho_0= \sqrt{k/\kappa}$ and $\kappa<0$, $k<0$. If we
take $\kappa=k=-1$ the soliton metric is then $e^{2r}
dt^2+dr^2+d\Omega_{-1}^2$ and is an expander.

The ${\dot \rho}\neq 0$ solutions climb the $V(\rho)$ potential well
until they reach one of the horizontal lines $k=0,\pm 1$, then turn
around and go to $V(\rho)\to -\infty$ as either $\rho\to\infty$ or
$\rho\searrow 0$. In the $\kappa<0$ case, it is possible to pass
between $\rho\searrow 0$ and $\rho\to\infty$ without encountering a
turning point (see the bottom curve of the Figure).

\begin{figure}
  \begin{center}
   \resizebox{8cm}{!}{\includegraphics{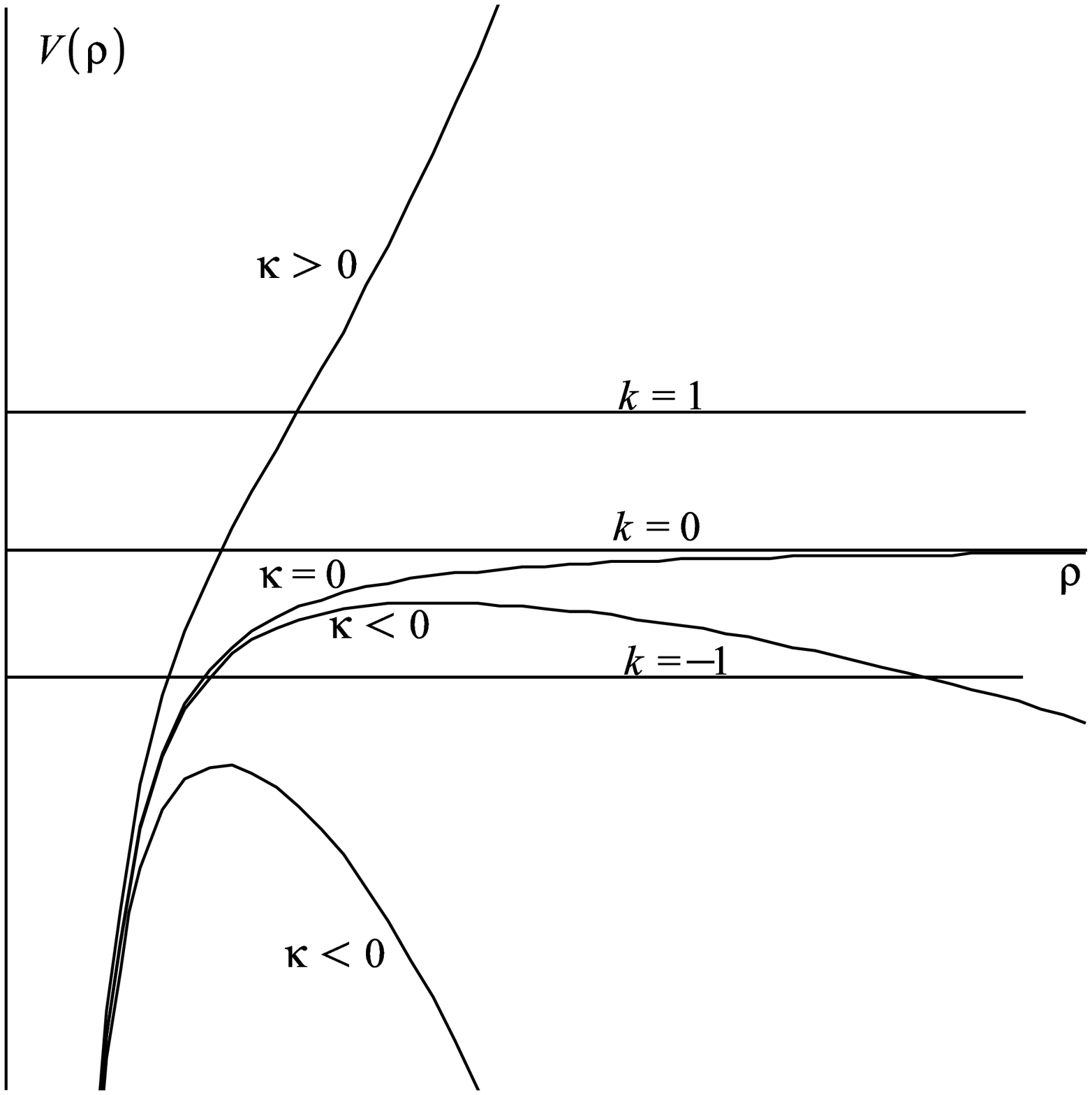}}
  \end{center}
\end{figure}

Integrating (\ref{eq3.9}) for small $\rho$, we see that the distance
coordinate $r$ is bounded as $\rho\searrow 0$, and ${\dot \rho} =
f'(r)$ blows up there. The sectional curvature in planes
perpendicular to $\frac{\partial}{\partial r}$ contains a
$(f'(r)/f(r))^2$ term and so also blows up. Thus, these metrics are
incomplete unless $\rho$ is bounded away from zero. The only
solutions with this property are those that move along curves such
as the higher of the two $\kappa<0$ curves in the figure. If a
solution starts out on this curve to the right of the rightmost
intersection point of this curve with the $k=-1$ line, then the
solution will move inward to the intersection point, then turn and
role back down the $\kappa<0$ curve, escaping to infinity. Such
solutions are complete. A concrete example is provided in section
{\ref{sec3.2.2}, but first we address the question posed at the
beginning of the section.

\subsubsection{Schwarzschild soliton}

\noindent The 4-dimensional (thus $n=3$) Schwarzschild metric gives
rise to a $\kappa=0$ local soliton. Writing the Schwarzschild metric
in the form of (\ref{eq3.1})
\begin{eqnarray}
d{\hat s}^2 &=& -\left ( 1-\frac{2m}{r} \right ) dt^2
+ \frac{dr^2}{\left ( 1-\frac{2m}{r} \right )}+r^2\left (
d\theta^2 + \sin^2 \theta d\phi^2 \right ) \nonumber\\
&=&-e^{2u}dt^2+e^{-2u}ds^2 \ , \label{eq3.10}\\
ds^2&=&dr^2+r^2\left ( 1-\frac{2m}{r} \right )\left (
d\theta^2 + \sin^2 \theta d\phi^2 \right )\ , \label{eq3.11}
\end{eqnarray}
we can read off $f$ and $u$. In particular,
\begin{eqnarray}
u&=&\frac12 \log \left ( 1-\frac{2m}{r} \right ) \label{eq3.12} \\
f(r)&=& \sqrt{r\left (r-2m\right )} \ . \label{eq3.13}
\end{eqnarray}
From (\ref{eq2.1}), the corresponding soliton is
\begin{equation}
ds^2=\left ( 1-\frac{2m}{r} \right )^{\sqrt{2}}dt^2
+dr^2+r\left ( r-2m \right )
\left ( d\theta^2+\sin^2 \theta d\phi^2\right ) \ .
\label{eq3.14}
\end{equation}
Since $\kappa=0$, we know the soliton metric will be incomplete
(e.g., by Remark 2.7). Indeed, from (\ref{eq3.12}) and
(\ref{eq2.11}), the soliton has scalar curvature
$-\frac{2m^2}{r^2(r-2m)^2}$, which diverges as $r\to 2m$.

Ivey has given examples of complete steady solitons in his PhD
thesis \cite{Ivey}. Our solitons are less general than Ivey's, since
in our case the same function $u$ appears in both the potential for
the vector field $X$ in the soliton equation (\ref{eq1.5}) and the
norm of the Killing field $\partial/\partial t$. The only
simultaneous solution of Ivey's soliton equations and ours in
dimension 4 is the local soliton (\ref{eq3.14}).\footnote
{One must take the $k=1$ case of Ivey's equations (cf Ch 5 of his
thesis \cite{Ivey}) and the $\kappa=0$ case of ours.}

If we Wick rotate both $t\mapsto it$ and $\phi\mapsto i\phi$ in
(\ref{eq3.10}), we are led to a different choice of $u$ and $f(r)$
and ultimately to the soliton
\begin{equation}
(r\sin\theta)^2\left [ (r\sin\theta)^{\sqrt{2}}d\phi^2
+\left ( 1-\frac{2m}{r} \right )dt^2
+\frac{dr^2}{\left ( 1-\frac{2m}{r} \right )}+r^2 d\theta^2 \right ]
\ , \label{eq3.15}
\end{equation}
which of course is also incomplete.

\subsubsection{A complete soliton} \label{sec3.2.2}

\noindent To obtain a complete soliton we must choose $\kappa<0$ and
$k=-1$. For illustrative purposes, we also choose the definite
values $\kappa=-1$ and $A=1/\sqrt{2}$. Then we can solve
(\ref{eq3.6}) to obtain
\begin{equation}
f^2(r)=1 + C e^{\pm \sqrt{2}r} \ . \label{eq3.16}
\end{equation}
Choose the plus sign and $C=1$ for definiteness. Then we get (using
(\ref{eq3.2}))
\begin{eqnarray}
ds^2&=&g_{ij}dx^idx^j = dr^2 +\left ( 1+e^{\sqrt{2}r} \right )
d\Omega^2_{k=-1}\ , \label{eq3.17}\\
\sqrt{2} u(r)&=&r-\frac{1}{\sqrt{2}}\log \left ( 1+e^{\sqrt{2}r} \right )
\ . \label{eq3.18}
\end{eqnarray}
The corresponding soliton is obtained by inserting these expressions
into (\ref{eq2.1}) to obtain
\begin{equation}
ds^2= \frac{e^{2r}}{\left ( 1+e^{\sqrt{2}r} \right
)^{\sqrt{2}}}\,dt^2+ dr^2 +\left ( 1+e^{\sqrt{2}r} \right )
d\Omega^2_{k=-1}\ . \label{eq3.19}
\end{equation}
Note that $r$ takes all real values. The sectional curvatures are
all negative except the sectional curvature in $t$--$r$ planes,
which is positive iff $r>-\frac{\log{2}}{2\sqrt{2}}$.

Since there is a compact hyperbolic 2-manifold for each integer
$g\ge 2$ and $3g-3$ distinct choices for the hyperbolic metric
$d\Omega_{k=-1}$ on each such manifold, (\ref{eq3.19}) represents a
countably infinite family of solitons. Also, we can, if we wish,
choose Lorentzian signature by Wick rotating $t\to it$, thereby
obtaining a Lorentzian Ricci soliton.

\subsection{$n=4$ Einstein-scalar solutions}

\noindent Above we regarded static vacuum metrics in $(n+1)$
dimensions as giving rise to a pair $(u,g_{ij})$ solving the static
Einstein equations an $n$-dimensional metric. A concrete example was
given by $4$-dimensional Schwarzschild metrics, which produced a
$3$-dimensional $g_{ij}$, leading back to a $4$-dimensional (local)
soliton metric. But we could also consider the trivial 5-dimensional
solitons arising from the pair $(u,g)=(0,g^{\rm Sch})$ comprised of
the zero function (or any constant function) and a 4-dimensional
Schwarzschild metric $g^{\rm Sch}$. Now this pair belongs within a
family of solutions of the static Einstein equations
\begin{eqnarray}
ds^2&=& \left ( 1-\frac{2m}{r} \right )^\delta d\tau^2+ \left (
1-\frac{2m}{r}\right )^{-\delta} dr^2\nonumber\\
&&+ r^2 \left (1-\frac{2m}{r}\right )^{1-\delta} d\Omega^{2}\ ,
\label{eq3.20} \\
u&=& \frac12 \sqrt{1-\delta} \log \left (1-\frac{2m}{r}\right )
\ , \label{eq3.21}
\end{eqnarray}
where $d\Omega^{2}$ is the standard round metric on
$\mathbb{S}^{n-2}$ and $\delta\in [0,1]$. These metrics are Wick
rotated solutions of the 4-dimensional Einstein-free scalar field
equations \cite{Fisher}. The higher dimensional generalization was
obtained in \cite{XZ}.

The local soliton corresponding to this solution is
\begin{equation}
\left ( 1-\frac{2m}{r}\right )^{3\sqrt{1-\delta}/2}dt^2+ds^2\ ,
\label{eq3.22}
\end{equation}
with $ds^2$ given by (\ref{eq3.20}). The volumes (i.e., surface
areas) of the $(n-2)$-spheres of constant-$r$ collapse to zero as
$r\searrow 2m$. Simultaneously, sectional curvatures of $g_{ij}$
diverge, as does the scalar field.\footnote
{Chase \cite{Chase} proved a no hair theorem showing that such
singularities occur even in the absence of the SO(3) rotational
symmetry. Chase used Lorentzian signature, prior to Wick rotation of
$\tau$, and assumed 4 dimensions, asymptotic flatness, and
$const/r+{\cal O}(1/r^2)$ fall-off for the radial part of the scalar
field (and one more power for the gradient). He found that, in the
presence of a nonzero scalar field, the Kretschmann scalar cannot
remain bounded on the domain of outer communications to the future
of an initial surface which may or may not contain a 2-sphere
apparent horizon. \label{fn5}}
By (\ref{eq2.13}), the 5-dimensional soliton constructed from this
metric inherits the divergence and is incomplete (as it must be,
since $\kappa=0$).

As well, there are known conformally flat solutions in 4 dimensions.
One class is due to Penney \cite{Penney}. Rotated to Euclidean
signature, they are given by
\begin{eqnarray}
g_{ij}&=& \left ( 1+a_i x^i\right ) \delta_{ij} \label{eq3.23}\\
u&=& \sqrt{\frac32}\, \log \left ( 1+a_i x^i\right ) \label{eq3.24}\ ,
\end{eqnarray}
where $C$ and $a_i$'s are arbitrary constants. Sectional curvatures
of $g_{ij}$ diverge on approach to the hyperplane $a_i x^i = -1$.
G\"{u}rses \cite{Gurses} found another class of such solutions,
again in 4 dimensions. These are
\begin{eqnarray}
g_{ij}&=& \left ( 1-\frac{k^2}{r^4}\right )\delta_{ij} \label{eq3.25}\\
u&=&\frac{\sqrt{3}}{2}\log \left \vert \frac{r^2-k}{r^2+k}\right
\vert+const. \label{eq3.26}
\end{eqnarray}
Here $r^2= \delta_{ij}\,x^i\,x^j$. Sectional curvatures diverge on
approach to the sphere $r=\sqrt{k}$. For both these examples, the
solitons constructed from them using (\ref{eq2.1}) also have
divergent sectional curvature (by (\ref{eq2.13})) and are
incomplete.

\section{Stationary metrics}

\noindent To close, we consider the more general class of stationary
metrics
\begin{equation}
ds^2=\pm e^{2\sqrt\frac{n-1}{n-2} u}\left ( dt+A_i dx^i \right )^2
+g_{ij}dx^idx^j \ , \label{eq4.1}
\end{equation}
where $\frac{\partial u}{\partial t}=0$, $\frac{\partial
A_i}{\partial t}=0$, and $\frac{\partial g_{ij}}{\partial t}=0$. The
$\pm$ sign will allow us to consider both signatures simultaneously.
The Ricci flow of this metric can be read off from equation (4.9) of
\cite{Lott}.

A reasonable approach is to mimic the procedure in the static case,
eliminating certain unwanted second derivative terms from the Ricci
flow by adding Lie derivative terms arising from the pullback via an
evolving diffeomorphism generated by vector field
$-\sqrt\frac{n-1}{n-2}\nabla u$ on the ``base manifold'' $(M,g)$. As
well, to eliminate an unwanted term in the evolution equation for
$A_i$, we perform the ``gauge transformation''
\begin{eqnarray}
B_i&:=&A_i-\nabla_i \Lambda\ , \label{eq4.2}\\
\frac{\partial \Lambda}{\partial \lambda} &=&
-\sqrt\frac{n-1}{n-2} \left ( B+\nabla \Lambda \right )\cdot \nabla u
\ , \label{eq4.3}
\end{eqnarray}
where we define
\begin{equation}
F_{ij}[A]:=\nabla_i A_j - \nabla_j A_i\ , \label{eq4.4}
\end{equation}
with $\nabla_i$ the Levi-Cevit\`a connection of $g_{ij}$. Of course
$F_{ij}[A]\equiv F_{ij}[B]\equiv F_{ij}$ is gauge invariant. Then we
obtain
\begin{eqnarray}
\frac{\partial u}{\partial \lambda}&=& \Delta u
\mp \frac{1}{4}\sqrt\frac{n-2}{n-1}
e^{2\sqrt\frac{n-1}{n-2}u} \vert F \vert^2 \ ,
\label{eq4.5} \\
\frac{\partial B_i}{\partial \lambda}&=& - \nabla^j F_{ij}
-2\sqrt\frac{n-1}{n-2}F_{ij}\nabla^j u
\ , \label{eq4.6} \\
\frac{\partial g_{ij}}{\partial \lambda}&=&
-2R_{ij} +2\left ( \frac{n-1}{n-2}\right )
\nabla_i u \nabla_j u \pm e^{2\sqrt\frac{n-1}{n-2}u}
g^{kl}F_{ik}F_{jl}\ . \label{eq4.7}
\end{eqnarray}
This system couples a scalar heat flow, a vector Yang-Mills flow,
and a generalization of Ricci flow, and can be made parabolic by
adding further gauge and diffeomorphism terms. Setting $F=0$, we
recover the static Einstein equations (\ref{eq1.8}, \ref{eq1.9}).

By way of comparison, the condition for the stationary metric
\begin{equation}
ds^2 = -e^{2u}\left ( dt+B_i dx^i\right )^2 +e^{-\frac{2u}{n-2}}g_{ij}dx^i dx^j
\label{eq4.8}
\end{equation}
to be Ricci flat is
\begin{equation}
{\bf S}_I:=\left ( \begin{array}{c} S_0\\ S_i\\ S_{ij}\end{array}
\right ) =0 \ , \label{eq4.9}
\end{equation}
where
\begin{eqnarray}
S_0&=&\Delta u +\frac14 e^{2\left (\frac{n-1}{n-2}\right )u} \vert F
\vert^2 \ ,
\label{eq4.10}\\
S_i&=&-\nabla^j F_{ij} - 2\left ( \frac{n-1}{n-2} \right ) F_{ij}\nabla^j u \ ,
\label{eq4.11}\\
S_{ij}&=&-2R_{ij}+2\left ( \frac{n-1}{n-2}
\right )\nabla_i u \nabla_j u\nonumber \\
&& -e^{2\left ( \frac{n-1}{n-2} \right )u}
\left [ F_{ik}F_j{}^k-\frac{1}{2(n-2)}g_{ij}\vert F \vert^2 \right ] \ .
\label{eq4.12}
\end{eqnarray}

Fixed points of the system (\ref{eq4.5}--\ref{eq4.7}) are obtained
by setting the time derivatives to zero so that the right-hand sides
vanish as well. The resulting equations differ from the system
(\ref{eq4.9}--\ref{eq4.12}) in the coefficients preceding some of
the $F$-terms. One consequence is that the fixed point condition for
the system (\ref{eq4.5}--\ref{eq4.7}) leads to an integrability
condition $e^{2\sqrt{\frac{n-1}{n-2}}u}|F|^2=const$ which does not
arise for the system (\ref{eq4.9}--\ref{eq4.12}).

If $F_{ij}\neq 0$, the fixed point condition for
(\ref{eq4.5}--\ref{eq4.7}) does not coincide with the condition that
the stationary metric (\ref{eq4.8}) be Ricci flat.

There are two alternative strategies. One is to apply a different
diffeomorphism and gauge choice than that used to obtain
(\ref{eq4.5}--\ref{eq4.7}).\footnote
{Presumably this would be combined with ``field redefinitions'' in
(\ref{eq4.1}); e.g., rewriting $A$ as $f(u)A$ with some suitably
chosen function $f$.}
Perhaps such a technique may produce fixed points for which ${\bf
S}=0$. However, a second technique is to study the flow
\begin{equation}
\frac{\partial}{\partial \lambda}\left ( \begin{array}{c} u\\ B_i\\ g_{ij}\end{array}
\right )={\bf S}\ . \label{eq4.13}
\end{equation}
This flow has the same well-posedness properties as
(\ref{eq4.5}--\ref{eq4.7}) and its fixed points $(u,B_i,g_{ij})$
make the stationary metric (\ref{eq4.8}) Ricci flat. It is an open
question whether, as we believe, (\ref{eq4.13}) is a genuinely new
flow, or whether a diffeomorphism can be found to bring
(\ref{eq4.13}) to the Hamilton-DeTurck form (\ref{eq1.2}).

\section*{Acknowledgments}
MMA would like to thank David Wiltshire for useful correspondence
and Hadi Salmasian for useful conversations. We thank Mike Anderson
for correspondence, RV Saraykar for correcting one of our citations,
and an anonymous referee for detailed suggestions that improved the
presentation. MMA was supported by a Pacific Institute for
Mathematical Sciences (PIMS) Post-Doctoral Fellowship. This work was
supported by an NSERC Discovery Grant awarded to EW.


\begin{thebibliography}{99}
\bibitem{Anderson1} MT Anderson, {\it Comparison Geometry}, MSRI Publications 30 (1997) 49.
\bibitem{Anderson2} MT Anderson, Geom Funct Anal 9 (1999) 855.
\bibitem{Bartnik} R Bartnik, {\it Proc Int Congress Math}, Beijing
    2002, vol 2, 231 [arXiv:math/0304259]; {\it Tsing Hua Lectures on
    Geometry and Analysis}, ed S-T Yau (International Press, Cambridge MA,
    USA, 1995) pp 5--27.
\bibitem{CZ}H-D Cao and X-P Zhu, Asian J Math 10 (2006) 169.
\bibitem{Chase} J E Chase,
    Commun Math Phys 19 (1970) 276.
\bibitem{CK} B Chow and D Knopf, {\emph{The Ricci Flow: An
    Introduction}}, Mathematical Surveys and Monographs vol 110
    (AMS, Providence, 2004).
\bibitem{Chow1} B Chow et al, {\emph{The Ricci Flow: Techniques and
    Applications: Geometric Aspects}}, Mathematical Surveys and
    Monographs vol 135 (AMS, Providence, 2007).
\bibitem{PTC} PT Chru\'sciel, Commun Math Phys 137 (1991) 289.
\bibitem{Fisher} IZ Fisher,
    Zh Eksp Teor Fiz 18 (1948) 636 [arXiv:gr-qc/9911008]; HA Buchdahl,
    Phys Rev 115 (1959) 1325; AI Janis, ET Newman and J Winicour,
    Phys Rev Lett 20 (1968) 878; M Wyman,
    Phys Rev D 24 (1981) 839; KS Virbhadra,
    Int J Mod Phys A12 (1997) 4831 [arXiv:gr-qc/9701021].
\bibitem{Geroch} R Geroch, Ann NY Acad Sci 224 (1973) 108.
\bibitem{Gurses} M G\"{u}rses,
    Phys Rev D15 (1977) 2731.
\bibitem{Huisken} G Huisken, private communication.
\bibitem{HI} G Huisken and T Ilmanen, J Diff Geom 59 (2001) 353.
\bibitem{Ivey} T Ivey, {\it On solitons of the Ricci flow}, PhD
    thesis (Duke University, 1992, unpublished); see also
    Proc Amer Math Soc 122 (1994) 241.
\bibitem{Lott} J Lott, preprint [arxiv:0711.4063].
\bibitem{JFL} Jun-Fang Li, unpublished notes.
\bibitem{Lich} A Lichnerowicz, {\it Th\'eories relativistes de la
    gravitation et de l'\'electromagn\'etisme} (Masson, Paris, 1955) p 138.
\bibitem{List} B List, {\it Evolution of an extended Ricci flow
    system}, PhD thesis 2005, Freie Universit\"at Berlin, unpublished.
\bibitem{MT2} J Morgan and G Tian, preprint [arxiv:0809.4040].
\bibitem{MT} J Morgan and G Tian, {\it Ricci flow and the Poincar\'e
    conjecture} (AMS, Providence, 2007) and references therein.
\bibitem{Penney} RV Penney,
    Phys Rev D14 (1976) 910.
\bibitem{Petersen} P Petersen, {\it Riemannian geometry} ($2^{\rm
    nd}$ edition, Springer, 2006) p 159.
\bibitem{Sullivan} D Sullivan,
    J Diff Geom 18 (1983) 723.
\bibitem{Tod} P Tod, Class Quantum Gravit 6 (1989) 1159.
\bibitem{XZ} BC Xanthopoulos and T Zannias,
    Phys Rev D40 (1989) 2564.

\end{thebibliography}
\end{document}